# Metastable inhomogeneous vortex configuration with non-uniform filling fraction inside a blind hole array patterned in a BSCCO single crystal and concentrating magnetic flux inside it

**Gorky Shaw[1,†], S. S. Banerjee[1,*], T. Tamegai[2] and Hermann Suderow[3]**

[1] Department of Physics, Indian Institute of Technology, Kanpur-208016, India
[2] Department of Applied Physics, The University of Tokyo, Hongo, Bunkyo-ku, Tokyo 113-8656, Japan
[3] Laboratorio de Bajas Temperaturas, Unidad Asociada UAM, CSIC, Instituto N. Cabrera, Condensed Matter Physics Center (IFIMAC), Facultad de Ciencias Universidad Autónoma de Madrid, E-28049 Madrid, Spain

*E-mail: satyajit@iitk.ac.in

[†] Present address: University of Liege (ULg), Department of Physics, Experimental Physics of Nanostructured Materials, Sart Tilman, B-4000, Belgium

**Abstract:** Using magneto-optical imaging technique, we map local magnetic field distribution inside a hexagonally ordered array of blind holes patterned in BSCCO single crystals. The nature of the spatial distribution of local magnetic field and shielding currents across the array reveals the presence of a non-uniform vortex configuration partially matched with the blind holes at sub-matching fields. We observe that the filling fraction is different in two different regions of the array. The mean vortex configuration within the array is described as a patchy vortex configuration with the patches having different mean filling fraction. The patchy nature of the vortex configuration is more pronounced at partial filling of the array at low fields while the configuration becomes more uniform with a unique filling fraction at higher fields. The metastable nature of this patchy vortex configuration is revealed by the application of magnetic field pulses of fixed height or individual pulses of varying height to the array. The metastability of the vortex configuration allows for a relatively easy way of producing flux reorganization and flux focusing effects within the blind hole array. Effect of the magnetic field pulses modifies the vortex configuration within the array and produces a uniform enhancement in the shielding current around the patterned array edges. The enhanced shielding current concentrates magnetic flux within the array by driving vortices away from the edges and towards the center of the array. The enhanced shielding current also prevents the uninhibited entry of vortices into the array. We propose that the metastable patchy vortex configuration within the blind hole array is due to a non-uniform pinning landscape leading to non-uniform filling of individual blind holes.

# 1. Introduction

In type-II superconductors dissipation is produced by vortex motion, which is suppressed by localizing vortices on pinning centers. Enhancing the pinning strength increases the threshold critical current density ($j_c$) of superconductors below which the current flow is dissipationless. Thus $j_c$ is a measure of the vortex pinning strength. Recent strategies to artificially enhance pinning involve making the vortex lattice commensurate with an artificially generated lattice (ordered array) of pins [1,2,3,4,5,6,7,8,9,10,11,12,13,14,15,16,17,18]. The vortex lattice is said to be commensurate with an ordered array of pins when the inter-vortex spacing ($a_0$), $a_0 \propto 1/\sqrt{B}$ (where $B$ is magnetic field) matches with the inter-pin spacing ($d$) (here we consider that each pin can accommodate only a single vortex). Thus the $B$ at which $a_0 = d$, defines a characteristic magnetic field, viz., the matching field, $B_\phi = \phi_0/d^2$ (where $\phi_0$ is the magnetic flux quantum). At $B = B_\phi$ the vortex density corresponds to an average of one vortex per pinning site. The presence of a periodic pinning array in a superconductor [7,19,20] modifies the magnetic field distribution [21,22,23] inside the material. Studies with periodic pins have usually used antidot lattices patterned in superconductors (antidots are through holes patterned in a superconductor) [1-11]. These studies have shown enhancements in vortex pinning at $B = nB_\phi$, where $n$ is an integer or fraction. Unlike antidots, there exists another alternate structure called blind hole array where the holes terminate inside the superconductor, viz., these are holes open only at one end. Studies suggest superconducting and vortex pinning properties associated with blind holes are quite distinct from those of superconductors with antidots [24,25]. A blind hole array produces a simply connected superconducting medium which is uniformly connected below the pins, as each hole in the array terminates inside the medium. A blind hole array has been considered as a prototype for a tuneable pinning landscape akin to Wigner crystals [26]. Theoretical studies [27,28,29] on superconductors with correlated array of holes (pins) show vortex configurations with different symmetries at sub-matching fields, along with novel dimer, trimer and giant vortex configurations possible within a blind hole array. While a large number of studies have investigated vortex configurations in antidot arrays, comparatively fewer studies exist on blind hole arrays [11-16,24,25]. In this paper using high sensitivity magneto-optical imaging technique we map the gradients in the local magnetic field across a high quality Bi$_2$Sr$_2$CaCu$_2$O$_8$ (BSCCO) single crystal patterned with a hexagonal array of blind nanoholes. Measurements at different applied field ($H$) and temperature ($T$) indicate the blind holes have a mean effective pinning strength of 70 K. We observe an enhancement in the pinning strength associated with partial filling of the blind hole array (matching effect) at sub matching field conditions. The mean filling fraction and the associated pinning strength is found to be different in two different regions inside the array. Thus the vortex configuration in the array is described as a patchy configuration, with different patches possessing different vortex filling fraction. The patchy non-uniform vortex configuration becomes more uniform at higher fields close to the matching field. The patchy vortex configuration of vortices at sub-matching fields is found to be metastable. At low fields we find the vortex configuration is quite amenable to reconfiguration by the application of a drive with a train of magnetic field pulses. The train of magnetic field pulses of constant pulse height produces an enhancement in the shielding currents uniformly around the array which helps in concentrating the magnetic flux towards the center of the array where they are trapped and vortices are driven away from the edges of the array. The uniformly large shielding current circulating around the edges of the array inhibits further easy entry of vortices into the array unless the magnetic field pulse height is increased. We propose the presence of a non-uniform vortex pinning landscape and the non-uniformly filled blind holes as possible sources of producing a metastable patchy vortex configuration within the blind hole array at low fields.

## 2. Experimental details and results

Our experiments are performed on a high-quality single crystal of $Bi_2Sr_2CaCu_2O_8$ (BSCCO) [30] of dimensions (0.8 × 0.5 × 0.03 mm$^3$) and $T_c$ = 90 K (cf. fig. 1(a)). The sample surface ('*ab*' crystal plane) was milled with Focused Ion Beam (FIB) machine (dual beam FEI make Nova 600 NanoLab) with a focused Ga ion beam (diameter ~ 7 nm) to create a hexagonal array of blind holes on the surface of the crystal, covering an area of ~ 39 μm × 44 μm. Figure 1(b) shows a magnified Scanning Electron Microscopic (SEM) image of a portion of the hexagonal array of holes. Mean diameter of each hole is 170 nm with mean center-to-center hole spacing (*d*) is 350 nm (BSCCO penetration depth $\lambda_{ab}$ ~ 200 nm for '*ab*' crystal orientation). The blind holes are open only on the top surface of the crystal with an average depth ~ 500 nm (< thickness of the crystal = 30 μm). For our hexagonal array of holes, $B_\phi$ = 195 Oe. We use high-quality high-$T_c$ single crystals to ensure weak intrinsic pinning in the pristine crystal.

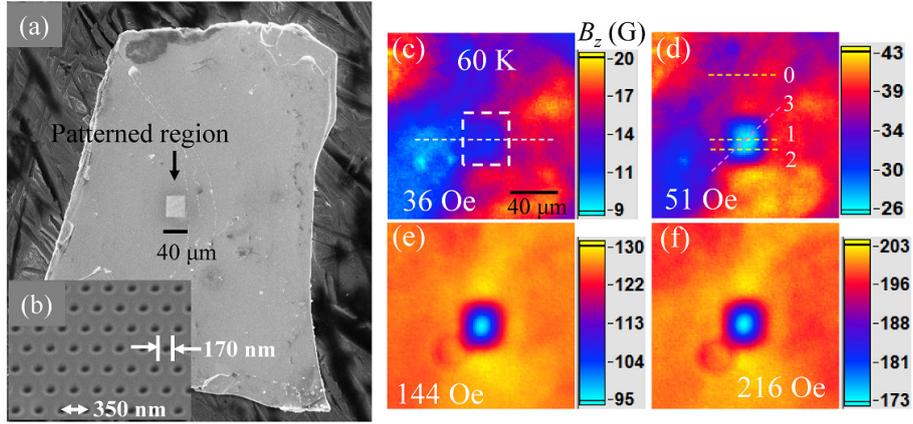

**Figure 1.** (Colour on-line) (a) SEM image of the crystal. The patterned region is the bright rectangular area (~ 39 μm × 44 μm) indicated by the arrow. (b) Zoomed SEM image of triangular array of holes with diameter 170 nm and center-to-center spacing of 350 nm. (c)-(f) $B_z(x, y)$ distribution in a 160 μm × 160 μm area around the patterned region at 60 K and *H* = 36, 51, 144 and 216 Oe, respectively. The scale bar beside each image represents the $B_z$ distribution (in Gauss) across the image. The patterned region is identified by the dashed square in (c).

We employ high-sensitivity magneto-optical (MO) imaging technique [30,31] to map the Faraday rotated light intensity distribution $I(x, y)$ across the surface of the superconductor. This technique enables us to image the distribution of the *z*-component of the local magnetic field $B_z(x, y)$ (as $I(x, y) \propto B_z(x, y)^2$). Figures 1(c) - 1(f) show at 60 K and different *H* the zero-field cooled $B_z(x, y)$ distribution in a 160 μm × 160 μm region in and around the patterned area. At 60 K and 36 Oe (fig. 1(c)), the colour contrast (or difference in $B_z$) across the edge of the patterned region (identified by the dashed square in fig. 1(c)) is found along three edges of the patterned region. Unlike the right edge, the left edge of the patterned region does not show any significant difference in $B_z$ value between inside and outside. Outside the patterned area, the differences in $B_z$ are due to small variations in pinning in the pristine crystal. At *H* = 51 Oe and higher (figs. 1(d) – 1(f)), the $B_z$ outside the patterned region becomes uniform as the orange shaded region begins to uniformly encircle the patterned region. With increasing *H* (figs. 1(d) – 1(f)) a strong contrast difference develops between the region with uniform orange shade outside and the bluish region inside the patterned area. The change in contrast across the edges of the patterned region indicates strong gradient develops in $B_z$ (viz., between inside and outside of the patterned region there is a slope in $B_z \propto J_c$) which in turn suggests the presence of stronger pinning inside the blind hole array relative to that outside.

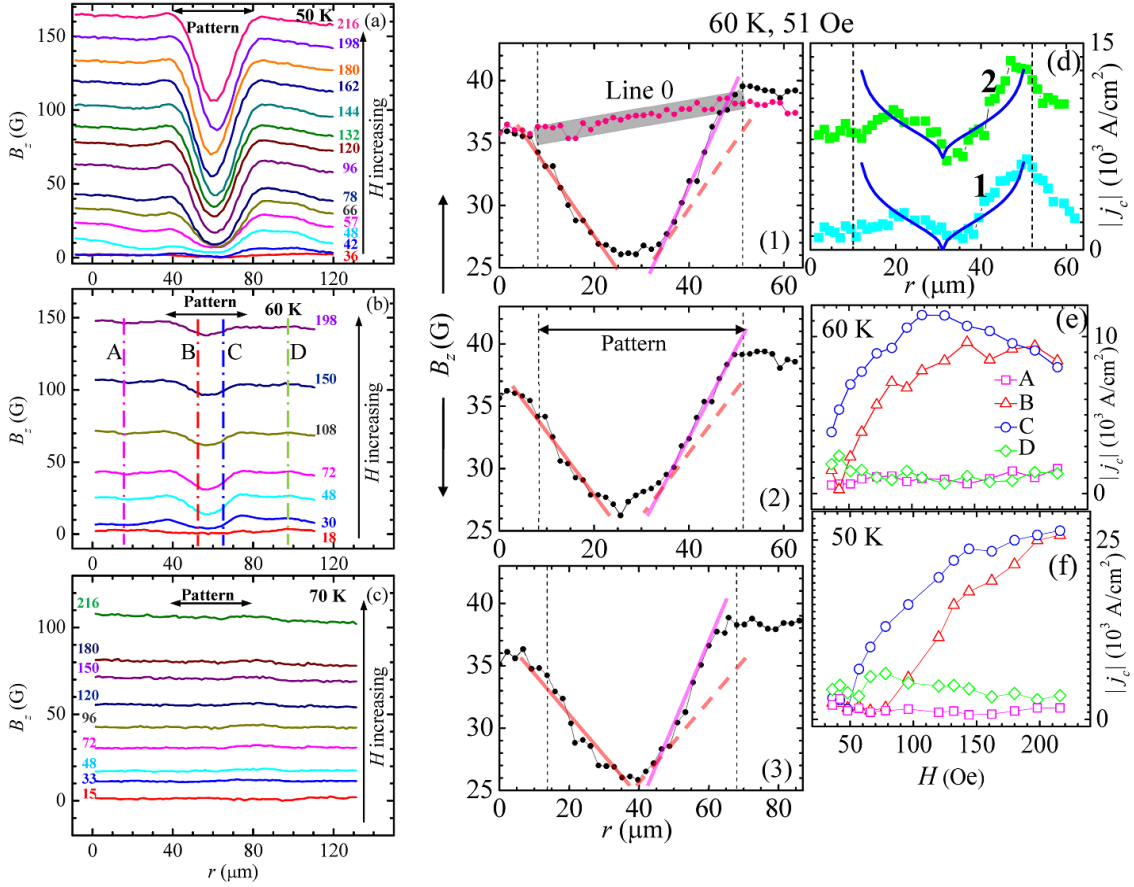

**Figure 2.** (Colour on-line) (a)-(c) $B_z(r)$ profiles at different $H$ measured across the line shown in fig. 1(c), at 50 K, 60 K, and 70 K respectively. Panels (1)-(3) show $B_z(r)$ determined along lines 1-3 marked in fig. 1(d). Panel (1) also shows the $B_z(r)$ profile measured along line labelled '0' in fig. 1(d). Solid lines (red and purple) approximately represent the gradient (slope) of the $B_z$ profiles near the right or left edges of the patterned region, respectively. The dashed (red) line is a mirror reflection of the solid (red) line drawn near the left edge. (d) $|j_c(r)|$ profiles across lines 1 and 2 (cf. fig. 1(d)). For clarity the $|j_c(r)|$ for line 2 is vertically offset by $6.7 \times 10^3$ A/cm$^2$. The vertical dashed lines in (1)-(3) and (d) identify the edges of the patterned region. The solid (blue) curves represent $|j_c(r)|$ calculated using conventional Bean critical state model (cf. text for details). (e)-(f) $|j_c(H)|$ profiles at four locations labelled A, B, C and D in (b) obtained from $|j_c(r)|$ profiles at different $H$, at 60 K and 50 K, respectively.

Figures 2(a) to 2(c) show $B_z(r)$ measured across the patterned region (viz., across a 120 μm line shown by a dashed line in fig. 1(c)) at different (increasing) $H$ for $T$ = 50 K, 60 K, and 70 K. It is clear from the line scans that inside the blind hole patterned region the $B_z(r)$ gradients are comparatively much larger than that outside. The contrast in the gradients between inside and outside the patterned region suggests that the pinning inside the blind hole array is stronger that that outside. Other studies have indicated that strong pinning is possible with blind holes [32]. From fig. 2(c) note that at 70 K, enhanced thermal fluctuations smear out the effective blind hole pinning inside the patterned region, and as a result the $B_z(r)$ profiles become flat and featureless, viz., there is no difference in the $B_z(r)$ between inside and outside the patterned region. Therefore the absence of gradients in the $B_z(r)$ at 70 K suggests that the average effective pinning strength in the blind hole array is ~ 70 K.

*2.1 Non-uniform gradients in $B_z$ profile within the blind hole array*

Across the unpatterned region of the crystal the $B_z(r)$ profile measured across line 0 in fig. 2(1) shows a weak gradient (weak as compared to that inside the patterned region) (cf. line 0 location in fig. 1(d)). The weak slope in $B_z(r)$ across line 0 suggests a weak pinning in the unpatterned regions of the sample, though it is much weaker as compared to that inside the patterned region. The standard deviation, $\sigma$ in $B_z(r)$ across line 0 is ~ 0.4 G. Panels 2(1) to 2(3) shows $B_z(r)$ profiles at 51 Oe and 60 K measured across the lines marked as 1 to 3 across the patterned region in fig. 1(d). In figs. 2(1) to 2(3) we draw solid lines (red and purple) to represent the average gradients (slope) in $B_z(r)$. In figs. 2(1) to 2(3) near the right edge of the patterned region we sketch a dashed (red) line which is a mirror reflection of the solid (red) line drawn near the left edge, viz., the dashed (red) and solid (red) lines in figs. 2(1) to 2(3) have the same magnitude of the slope. The difference in slope of dashed (red) line and the solid (purple) line shows the gradient in $B_z(r)$ is non-uniform inside the patterned region. The unequal magnitude of the slopes in $B_z(r)$ should correspond to differences in shielding currents inside the patterned region.

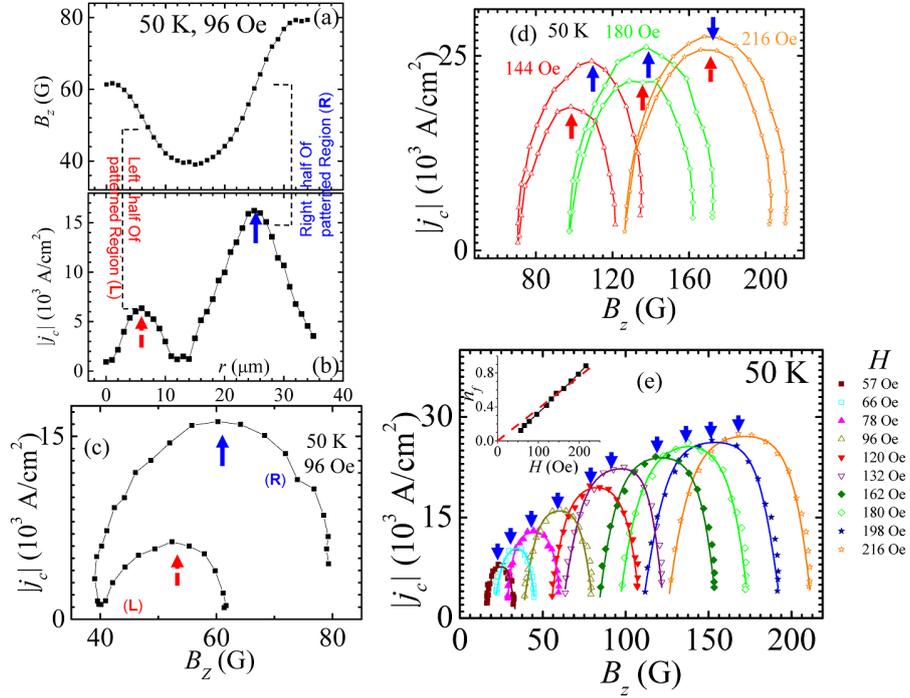

**Figure 3.** (Colour on-line) (a) and (b) $B_z(r)$ and $|j_c(r)|$ profiles, inside the patterned region at 50 K, 96 Oe extracted from data along a line 1 in fig. 1(d). (c) $|j_c(B_z)|$ at 50 K, 96 Oe (cf. text for details). (d) $|j_c(B_z)|$ at 50 K at different $H$ as indicated. In (c) and (d), the red and blue arrows indicate the maxima in $|j_c|$ corresponding to the left (L) and right (R) halves of the patterned region, respectively. (e) $|j_c(B_z)|$ only for the right side of patterned region at 50 K at different $H$. Inset of (e) shows variation of the filling fraction $n_f$ vs $H$ at 50 K. The red dashed line is a straight line fit through the origin.

By numerically inverting [33,34] $B_z(x, y)$, we determine the absolute value of the shielding current distribution $|j_c(r)|$ inside the patterned region. Figure 2(d) shows $|j_c(r)|$ determined across lines 1 and 2 (fig. 1(c)) (for clarity the $|j_c(r)|$ for line 2 is artificially offset by $6.7 \times 10^3$ A/cm$^2$). In fig. 2(d) the smooth (blue) curves represent $|j_c(r)|$ calculated using modified Bean model [22,23]. The calculation is restricted to within the patterned array and in the calculation we consider the array as a conventional superconducting medium without blind holes. Note from fig. 2(d) that the $|j_c(r)|$ on the left and right half of the patterned region is non-uniform. The non-uniformity in $|j_c(r)|$ reconfirms the non-uniformity in gradients of $B_z(r)$. Figures 2(e) and 2(f) show $|j_c|$ versus $H$ at four locations indicated in Fig. 2(b) at 60 K and 50 K, respectively. The $|j_c(H)|$ values are measured at locations B and C at 60 K and 50 K (see B and C locations in fig. 2(b)). Similarly figs. 2(e) and 2(f) shows $|j_c(H)|$ (pink and green curves) measured at A and D outside the patterned region (see locations in fig. 2(b)). At different $H$ the difference in $|j_c|$ between the red (left half) and blue (right half) curves inside the patterned region, viz., different in $|j_c|$ at B and C, is clearly identifiable. The non-uniformity in gradients of $B_z$ inside the patterned region diminishes only at $H$ values close to $B_\phi$ = 195 Oe, where the B and C curves approach each other in figs. 2(e) and 2(f). At lower $T$ = 50 K (fig. 2(f)) the non-uniformity in $|j_c(H)|$ inside the patterned region is much larger than that outside. The small difference in $|j_c(H)|$ values on left (A) and right (D) regions outside the patterned array at 50 K suggests the intrinsic pinning of the pristine crystal is also not completely uniform.

*2.2 Evidence of a partially filled vortex configuration and variation in the filling fraction across the blind hole array*

Using the $B_z(r)$ and $|j_c(r)|$ profiles determined at 50 K at different $H$ inside the patterned region (e.g., see figs. 3(a) and 3(b)) we determine the behaviour of $|j_c|$ versus $B_z$ inside the patterned region. Figures 3(c) and 3(d) show the $|j_c(B_z)|$ behaviour at 50 K inside the patterned region at different $H$. The $|j_c(B_z)|$ in fig. 3(c) shows that on the right half (R) of the patterned region, the maximum $|j_c|$ is reached at $B_z \sim 60$ G (at a location which is inside the patterned region, cf. fig. 3(a)). This maximum in $|j_c|$ corresponds to a vortex configuration partially filling the blind hole array where the pinning strength and hence $|j_c|$ peaks. The $B_z = 60$ G (arrow location in fig. 3(c)) is associated with a filling fraction $n_f = B_z/B_\phi$ of $\sim 0.31$. At the same $H$ and $T$, fig. 3(c) shows that in the left half of the patterned region, $n_f$ is different (0.27) and the maximum $|j_c|$ value is down by about 50%. Therefore we observe a difference in $n_f$ between the left and right half of the patterned region along with difference in $|j_c|$. Figure 3(d) shows that at $H$ = 144 Oe, a small difference in $n_f$ and $|j_c|$ maxima values between the right and left edges of the patterned region persists. With further increase in $H$ the difference in $n_f$ and peak $|j_c|$ between the left and right sides of the patterned region also reduces. At $H$ = 216 Oe, $n_f$ = 0.87 is identical on both sides of the patterned region and the peak in $|j_c|$ are also nearly similar. Figure 3(e) shows $|j_c(B_z)|$ at different $H$ at 50 K for only the right half of the patterned region. Inset of fig. 3(e) shows that as $H$ is increased the filling fraction $n_f = B_z/B_\phi$ increases monotonically, approaching unity (viz., one vortex per hole). Figure 3(e) also shows that the maxima peak in $|j_c|$ and hence the peak in pinning strength at partial filling increases with $n_f$ (located by arrows). Since increasing $H$ results in a larger fraction of vortices getting pinned on blind holes sites hence the increase in pinning with $n_f$ is expected. Similar behaviour for $|j_c(B_z)|$ is also found at 60 K (data not shown). We have already noted that the filling fraction ($n_f$) of the vortex configuration which is matched with the blind hole array is not uniform across the blind array, especially at low $H$. The above non-uniformity in $n_f$ values seen between the left and right sides of the patterned region suggests that the vortex configuration which is partially matched with the blind holes in the array is like patches of vortex configuration where each patch has a different $n_f$. One patch resides close to the right edge of the patterned region and another one on the left.

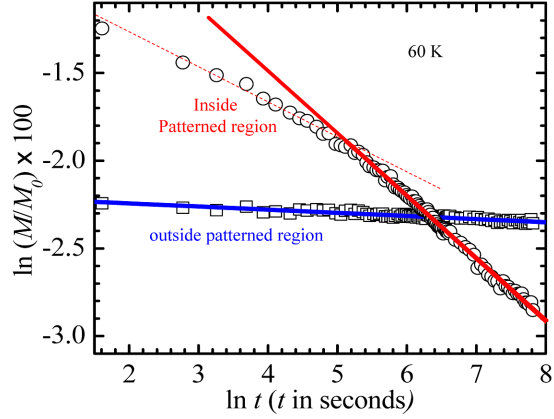

**Figure 4.** (Colour on-line) Magnetization relaxation response ln($M/M_0$) vs. ln $t$ inside the patterned (circles) and unpatterned (squares) regions as determined from MO images at 60 K (cf. text for details).

*2.3 Behaviour of local magnetization relaxation rate within the nanopatterned region*

Figure 4 shows the relaxation of the local remnant magnetization [30] ($M_{rem}$ vs $t$) measured within the patterned region. MO images are captured at intervals, $t$ = 200 ms to determine the $B_z(t)$ averaged over the patterned region and we determine the relaxation of $M_{rem}(t) = B_z(t) - H$. The remnant state is produced by field cooling the sample in a field of 150 Oe from 100 K down to 60 K and then switching off $H$. Figure 4 shows $ln\left(M_{rem}(t)/M(t=0)\right)$ vs. $ln(t)$ at 60 K. The relaxation rate of $M_{rem}(t)$ is related to thermally activated hopping of vortices over the pinning potential barriers [35,36], which in turn gives a measure the effective strength of the pinning potential. Figure 4 shows weak $M_{rem}$ relaxation outside the patterned regions (squares) indicating weak intrinsic pinning in the pristine region of the crystal. In fig. 4 the two linear fits (solid (red) and dashed (red) line) with a change in slope indicate that at shorter time scales the average relaxation rate is slower than that at longer times. The above suggests a distribution in the pinning strength of the vortex configuration which partially fills the blind hole array.

*2.4 Effect of magnetic field pulse of varying pulse height on vortex distribution within the array*

We now study the effect of a non-contact drive on $n_f$ inside the blind hole array. For applying a non-contact drive on the vortex configuration within the array we apply magnetic field pulses in the following way: The crystal is first zero-field cooled to below $T_c$ and subsequently a field $H_1$ is applied and MO images are captured at $H_1$. The $H$ is then increased *abruptly* to a higher value $H_2$ ($H_1 + \delta H$, where we have used a pulse of height $\delta H$ of 12 Oe). The $H$ is held constant at $H_2$ for 20 secs, after which $H$ is reduced back to $H_1$ and MO images are captured at $H_1$. The above protocol is repeated by increasing the pulse height from $\delta H$ to $2\delta H$, $3\delta H$, $4\delta H$… and so on (cf. schematic in Figure 5(g)). Note irrespective of size of $\delta H$, the MO images before and after the pulse are always captured when $H$ returns back to $H_1$, they are not captured at $H_2$. In our measurements $H_1$ is set at 36 Oe as, at 60 K, 36 Oe is just above the bulk penetration field of the crystal ~ 30 Oe (estimated from $M(H)$ curves at 60 K (not shown) similar to those shown in Ref. 13 at 50 K). We wanted to check if $n_f$ inside the patterned region increases easily with the application of magnetic field pulses of increasing height.

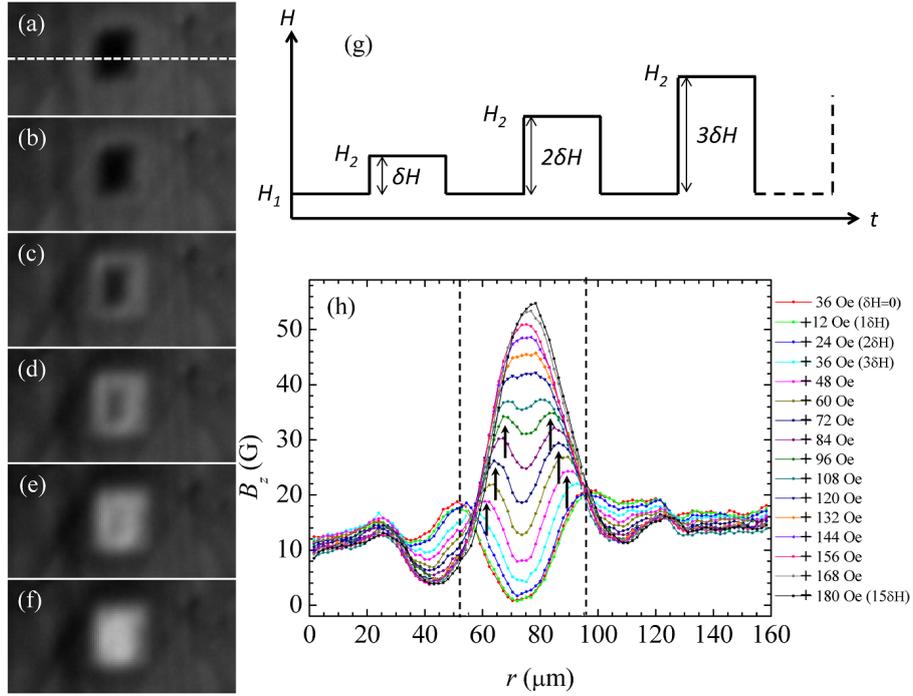

**Figure 5.** (Colour on-line) (a) MO image obtained at 60 K, 36 Oe before applying a higher field pulse. (b)-(f) MO images at 36 Oe obtained after one step change of field from $H_1 = 36$ Oe to $H_2 = 60, 84, 108, 132,$ and $156$ Oe, respectively (viz. pulse height $k\delta H = 24, 48, 72, 96,$ and $120$ Oe, respectively), and back to $H_1 = 36$ Oe. (g) Schematic showing the sequence of application of successive increasing field pulses. (h) $B_z(r)$ profiles across the line shown in (a) at 60 K, 36 Oe after application of successive field pulses. The pulse height $k\delta H$ corresponding to each profile is indicated alongside (cf. text for details).

Figure 5(a) shows zoomed in image of the patterned region obtained at $H_1 = 36$ Oe before applying any field pulse. Figures 5(b) to 5(f) show the same region at 36 Oe after applying field pulses $H_2 = H_1 + k\delta H = 60$ ($k = 2$), 84 ($k = 4$), 108 ($k = 6$), 132, ($k = 8$) and 156 ($k = 10$) Oe, respectively. Figure 5(h) shows $B_z(r)$ profiles measured across the line in fig. 5(a) after application of successive field pulses. Clearly from the images in (a)-(f), as the pulse height is increased, the contrast of the patterned region changes from a darker to brighter contrast as compared to the surrounding unpatterned regions, and the brightened regions appear to be moving from the edges and towards the center of the array. This is also seen in the line scans in fig. 5(h), where the peak in $B_z$ (located by arrows) progressively moves inside the patterned region from the edges of the array with the application of pulses of increasing height. Thus it appears that with every succesive magnetic field pulse of increasing height magnetic flux is swept from the edges to the center of the array. This sweeping of flux from the edges of the patterned region towards the center of the array also results in a depletion of vortices near the edges of the array where we find a reduction in $B_z$ at the edges of the array (cf. fig. 5(h)). This effect naturally leads to a reversal and enhancement in the $B_z$ gradient across the array, viz., compare the $B_z$ gradients inside the patterned region at $H = H_1$ and at $H = H_1 + 15\delta H$ in fig. 5(h) and note that the gradient in $B_z$ for field pulse $H = H_1 + 15\delta H$ is opposite in sign and larger compared to that at $H = H_1$. Note that we are always measuring $B_z$ when $H$ returns to $H_1$ after the field pulse. However the difference in the profiles suggests significant hysteretic response.

One may consider that the above effect is a result of additional vortices introduced in the patterned region by the field pulse ($k\delta H$), and these additional vortices are pushed deeper into the array with successive pulses. These vortices get trapped inside the array and collect near the center of the array and this would explain the $B_z$ increasing in the center of the array and the gradient inside the array reversing in sign. If this scenario were to be correct, then $n_f$ inside the patterned region would have increased quite rapidly with increasing pulse height. However we observe only a modest increase in $B_z$ in the center of the array when a large $k\delta H$ pulse is applied. Figure 5(h) shows that for $k = 15$, viz., $H_2 = 216$ Oe (= 36 Oe + 15 × 12 Oe) the mean $B_z$ at the center of the array is about 54 G which corresponds to an $n_f \sim 0.28$, while the size of the field pulse $15\delta H = 180$ Oe is of the order of the matching field of the array (195 G). It is worthwhile comparing $n_f \sim 0.28$ obtained with these pulsing experiments where the peak value of 216 Oe is reached momentarily, while $n_f \sim 0.87$ obtained at a constant dc field of 216 Oe (cf. fig. 3(d)) when the $H$ was applied by ramping the magnetic field slowly from 0 Oe towards the target field of 216 Oe. It appears that the effect of applying magnetic field slowly or fast affects the blind hole patterned region differently. The increase in $B_z$ inside the patterned region in the above experiment doesn't seem to be explained by trapping of flux inside the patterned region due to cycling of magnetic fields. From these measurements it appears that although there exists a sizeable number of sites available for occupation by the vortices they are not easily accessible with the application of the magnetic field pulses. The mechanism does not appear to be one of pushing additional vortices into the blind hole array by increasing the magnetic field and trapping them on the blind hole pins. The above measurements suggest the metastability of the vortex configuration within the array. The metastable nature of the vortex configuration in turn allows for relatively easy ways of reconfiguring vortices inside the patterned array as we shall show below.

*2.5 Effect of multiple field pulses of fixed height on vortex distribution within the array*

We investigate the above effect more closely. In this experiment, instead of applying single pulses of increasing height (as in the above experiment) we keep the height of each pulse fixed and apply the pulses multiple number of times (*n*) and observe the effects on the flux distribution in and around the patterned region. Does the application of multiple pulses of fixed height lead to an enhancement in $n_f$ at the center of the array? In this experiment a single pulse cycle ($n = 1$) corresponds to $H_1 \rightarrow H_2 \rightarrow H_1$. Higher field pulse cycles correspond to repeating $H_1 \rightarrow H_2 \rightarrow H_1$, *n* no. of times ($n = 0$ to 100), with $H_2$ kept fixed at 108 Oe, and MO images are captured only after the *H* returns to $H_1 = 36$ Oe after *n* cycles. Figures 6(a) to (c) shows zoomed in images of the patterned region obtained at $H_1 = 36$ Oe for $n = 0$, 1 and 100, respectively ($H_2 = 108$ Oe). The three-dimensional maps adjoining figs. 6(a) to (c) show that inside the patterned region, while there is a valley-like feature in the MO intensity $I(x, y)$ ( $\equiv B_z(x, y)$) for $n = 0$, it transforms into a dome-shaped feature for $n = 100$ as flux is driven from the edges of the patterned region towards the center where they are trapped. Figure 6(d) shows $B_z(r)$ profiles measured across the patterned region (along white line in fig. 6(a)), for $n = 0$, 1, 5, 20 and 100. Note that in fig. 6(d) for $n = 0$ at 36 Oe, the $B_z(r)$ profile has the same non-uniformity (between the red dashed line and the purple solid line) as noted in figs. 2(1) to 2(3). The profile near the left edge of the patterned region has a lower slope compared to the right edge. With $n = 1$ pulses, the concave like $B_z$ profile in fig. 6(d) transforms into a convex profile (red curve). In the MO image of fig. 6(b) this behaviour shows up as orange MO contrast (regions with high $B_z$) entering inside the patterned region from the edges compared to fig. 6(a) with a bluish core near the array center for $n = 0$. With successive pulses, for example by $n = 20$ the $B_z(r)$ near the center of the patterned region has significantly enhanced (cf. fig. 6(d)). Fig. 6(c) shows for $n = 100$, $B_z \sim 32$ Oe (orange shade) is concentrated in the center of the patterned region.

Note that between $n = 20$ to 100 cycles the $B_z$ profile remains almost unaltered, suggesting a maximum trapping of flux has been achieved with this procedure. The peak in $B_z$ saturating to ~ 32 G with $n > 20$ implies a peak filling fraction of $n_f \sim 0.16$ in the center of array and $n_f$ doesn't increase any further. In principle one would have expected that with $H_2 = 108$ Oe applied during the pulses, $n_f$ should reach close to a value of $n_f \sim 108/195 \sim 0.55$. Figure 6(e) shows the behavior of $|j_c(r)|$ profile [33,34] determined across the patterned region for $n = 0$ and $n = 100$ cycles. Like fig. 2(d), fig. 6(e) shows the non-uniformity in $|j_c(r)|$ profile for $n = 0$. With the application of pulses ($n = 100$) the $|j_c|$ near the left edge increases significantly compared to the right edge. Thus it appears that with applying field pulses the patterned array responds by uniformly enhancing the shielding currents circulating around the patterned regions. These enhanced shielding currents play an important role in preventing the enhancement of $n_f$ within the array when magnetic field pulses are applied. This effect doesn't happen when the $H$ is increased slowly. The above would help explain the difference in $n_f$ when $H$ is increased slowly compared to when it is increased with a field pulse, as noted in the earlier section.

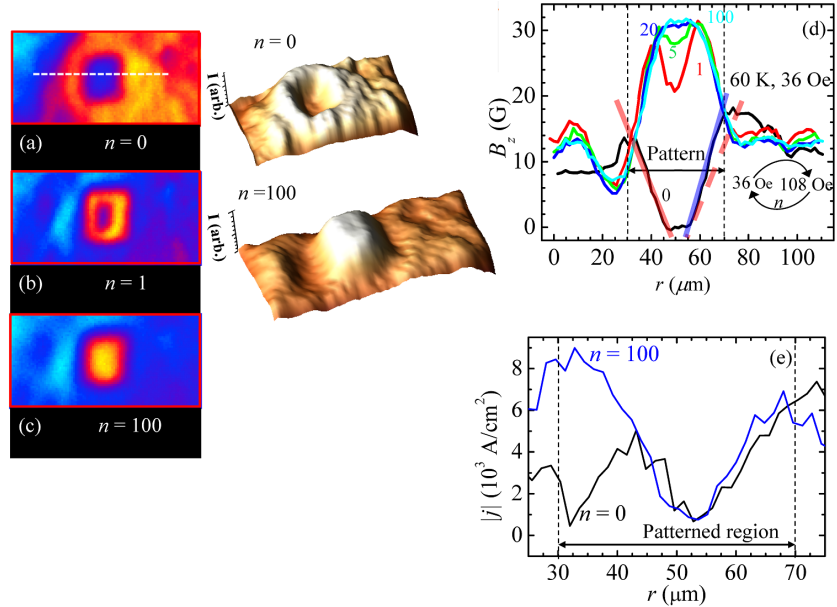

**Figure 6.** (Colour on-line) (a) MO image obtained at 60 K, 36 Oe ($n = 0$). (b) MO image at 36 Oe obtained after one step change ($n = 1$) of field from $H_1 = 36$ Oe to $H_2 = 108$ Oe back to $H_1 = 36$ Oe. (c) MO image at 36 Oe, with $n = 100$. The colour code represents the $B_z$ variation shown in fig. 6(d). The adjoining figures are three dimensional maps of the MO intensity distribution shown in (a) and (c). (d) $B_z(r)$ behaviour across the line shown in (a) for $n = 0$ (black), 1 (red), 5 (green), 20 (blue) and 100 (cyan). (e) Average current density distribution $|j(r)|$ across the patterned region for $n = 0$ (black) and 100 (blue) (cf. text for details).

## 3. Discussion of results

The difference in the arrow locations (red and blue arrows) in figs. 3(c) and 3(d) had indicated that $n_f$ is not uniform in different regions of the patterned array at low $H$. As has already been mentioned, this observation suggests the existence of patches of vortex configuration within the array where each patch has a different filling fraction, $n_f$, i.e., different patches have different $n_f$. The patchy nature of the vortex configuration is more pronounced at partial matching conditions, viz., at low field values which are less than the matching field. The patchy configuration becomes more uniform as the magnetic field approaches $B_\phi$, viz., in fig. 3(d) the location of the red and blue arrows begin to nearly coincide as $H$ is increased. Recall from fig. 2(f) that at 50 K we had found the right side outside the patterned region (viz., location D) has slightly higher $|j_c|$ compared to the left side outside the patterned region (viz., location A). We believe that the intrinsic pinning landscape in the pristine crystal is not uniform and in this non - uniform pinning background in the crystal affects the uniformity of blind hole pinning. Due to a superposition of the pinning landscapes, viz., that present in the pristine crystal and the blind hole pinning, the effective blind hole pinning landscape is not uniform. Due to this non - uniformity of the effective blind hole pinning landscape we observe the non-uniform gradients in $B_z(r)$ and $|j_c(r)|$ inside the patterned region in fig. 2. Matching of vortices with the blind holes with a particular $n_f$ inside the patch depends on a balance between inter-vortex and vortex-pin interactions [27,28] in the patch. As there exists small differences in the effective blind hole pinning within the array, patches with different $n_f$ begin to appear in different regions of the array. The presence of vortex states inside the blind hole array with different pinning strengths was also evident from different magnetization relaxation rates measured inside the patterned region at 60 K in fig. 4. We investigate the possibility that the non-uniformity in the effective underlying pinning landscape within the blind hole pinning array may be a possible source of producing the metastable patchy vortex configuration. In fig. 2(l), a comparison shows that the gradient in $B_z$ inside the patterned region (that is for the orange line in fig. 2(l)) is about ten times than that in the unpatterned region of the sample. Therefore it seems that the pinning in the underlying pristine unpatterned crystal is far weaker than the pinning of the blind holes. In such circumstances the extent of non-uniformity produced in the effective pinning of the blind holes by intrinsic (pristine crystal) pinning does not appear to be a very significant effect. Hence it seems unlikely that the effect of the non-uniformity of the underlying effective intrinsic pinning landscape on the pinning of the holes could be the only possible source for generating the metastable response inside the array. Another likely source of metastability at sub-matching fields could be the blind hole array itself. Even when the intrinsic pinning in the crystal is uniform vortices approaching from the edge of the patterned region towards its center would always experience a barrier due to vortices already present inside blind holes. The presence of these barriers would prevent the vortex state system from easily accessing the theoretically predicted ordered ground state vortex configuration composed of an ordered collection of multiple vortices pinned on each blind hole pin [27]. Experimentally it has been shown that unlike antidots, blind holes do not capture equal number of vortices [24,25]. Below the matching field, a situation could arise where multiple vortices are pinned at some blind holes while other holes remain unfilled. In such cases, patches with different $n_f$ could appear in different regions within the patterned area depending on the local balance between inter-vortex and vortex-pin interactions. Such a situation would be encountered even if the effective blind hole pinning is considered identical for all holes. Recent experiments show by direct observation that a one dimensional (1D) periodic thickness modulation in a thin film can provide a scale invariant random potential, when the vortex lattice is oriented at an arbitrary angle with the periodic modulation [37]. At high magnetic fields, detailed observations including many thousands of vortices show that the scale invariant random potential disorders the vortex lattice. The 1D periodic modulation does not provide any 1D pinning features. Instead, it leads to disorder through the misorientation of the vortex lattice with respect to the 1D modulation. Here, we are investigating a two dimensional (2D) periodic pinning array at sub matching fields, which is quite a different situation. Nevertheless, we can speculate that incomplete filling within the 2D blind hole array, or misorientation between both lattices, modifies the collective pinning landscape within the 2D array. The observed patchy nature of the vortex configuration within the patterned region could arise from interactions between the pinning and the vortex lattices. At low fields, the barrier for the penetrating vortices can lead to additional disorder, enhancing the non-uniformity in the pinning landscape. Stronger inter-vortex interactions at high magnetic fields would reduce the degree of non-uniformity of vortices filling the blind holes. At higher fields, close to the matching field, both lattices would lock, reducing the non-uniformity in vortex arrangements. In fact we observe the non-uniformity in $n_f$ decreases with increasing $H$ (cf. fig. 3(d)). Some of the above features pertaining to nature of the pinning landscape within a 2D periodic pinning array of blind holes is worth investigating for the future.

We have explored the nature of metastability of the patchy vortex configuration within the array by cycling the magnetic field. Earlier studies on dynamic reordering effects [38] had investigated the effects of drive on a uniformly pinned vortex medium. Instead here we are studying the effect of drive on non-uniform vortex state. To explain the details of the results reported in fig. 6, we suggest the following: the magnetic field pulse generates a shielding current $j_s$. As different regions have patches of vortices with different effective pinning, therefore the $j_s$ exceeds $j_c$ locally for a weakly pinned patch near one side the array, while $j_s$ would be smaller than $j_c$ in another stronger pinning patch within the array. Thus vortices in the weakly pinned patch near the left half of the patterned area are depinned and driven deeper into the array by the pulse while those in the strongly pinned regions near the right half remain unaffected. Thus the asymmetry in pinning around the patterned region may result in a directional flow of vortices into the patterned region, viz., flow from weaker pinned patches inside the array towards the center of the array. In fact fig. 6(e) shows that the $|j_c|$ value near the right edge of the patterned region is not affected much in comparison to the left edge where there is a significant increase in $|j_c|$ value as $n$ changes from 0 to 100. The drive induced by the field pulses drives the weakly pinned patch of vortices towards the center of the array where they get pinned resulting in an enhancement in $B_z$ gradient near the left edge. By driving the metastable patchy vortex configuration, the magnetic field pulses enhance the $|j_c|$ circulating near the left edge and consequently the non-uniformity in $j_c$ between two edges (as seen in fig. 2(d) and figs. 3(b) and 3(c)) diminishes (cf. fig. 6(e)). With multiple field pulses the shielding current circulating around the patterned array becomes uniformly large. This large shielding current circulating around the patterned array edges drives vortices away from the edges and compresses [39,40,41] the magnetic flux towards the center of the array. The presence of uniformly enhanced shielding currents circulating around the patterned array edges shields the interior of the array from continuous entry of flux as $H$ is switched between 36 Oe and 108 Oe, due to which $n_f$ saturates to about 0.16 as noted earlier. Increasing the height of the field pulse is not sufficient to make additional vortices enter into the patterned array, due to which we do not observe the increase the $n_f$ with the pulse height. The presence of large shielding currents around the patterned array prevents any large increase in $n_f$ within the array. If $H$ is increased gradually as in figs. 2 and 3 to reach the target $H$ at which measurements are performed then the shielding currents are much weaker and allow for easier flux entry into the patterned region. Thus the metastability and patchy nature of the vortex configuration allows for relatively easy ways of producing flux reorganization and flux focusing effects within the blind hole array. To the best of our knowledge, such flux reorganization effects with field pulses within a nanopatterned array created in a single crystal hasn't been shown before either for blind holes or antidot lattices patterned in superconductors.

Metastable nature of these systems could be further explored by other means, for example, by studying the effect of thermal cycling on the vortex state that is reached with $n$ = 100 in fig. 6(d) or by repeated fast switching of driving current, which in turn would correspond to subjecting the vortex state to a high effective shaking temperature [38]. The high shaking temperature would allow the vortex state to reach a lower energy state. If the metastability arises solely due to the blind hole array, the consequent flux reorganization could lead the system to gradually approach a more ordered, and eventually, a completely ordered state. On the other hand, if the metastability arises due to an interplay between intrinsic pinning in the sample and the blind hole array, the system might continuously evolve into different metastable configurations without being able to approach a more ordered state. In this regard recall our observation of a dome shaped $B_z$ profile developed across the blind hole array with $n$ = 100 field cycling (see fig. 6, bottom figure in second column of figures). It is known that in bulk single crystals of high-$T_c$ superconductors with ordered vortex configuration where the bulk pinning is negligible, the barriers from the sample edges result in a dome shaped $B_z$ profile [39,42]. Thus the observed dome shaped $B_z$ profile with $n$ = 100 field cycling suggests the vortex state within the array approaches a uniform distribution with repeated field cycling rather than becoming more metastable.

**4. Conclusion**

A superposition of a non-uniform intrinsic pinning landscape and uniform blind hole pinning results in an non-uniform pinning landscape within the blind hole array. We emphasize that although one may fabricate a nanopatterned structure (blind hole array in our case) which appears geometrically symmetric, there is an underlying non-uniformity in the effective pinning which leads to non-uniform field gradients inside the patterned region. The non-uniform pinning landscape within the blind hole array results in a patchy like vortex configuration of the array where each patch has a different filling fraction. Such a configuration is found to be metastable. We observe application of multiple field pulses of fixed height on such a configuration enhances the shielding current uniformly around the patterned array and consequently compresses the trapped flux towards the center of the array. The enhanced shield current prevents further entry of flux into the array unless the height of the pulse is increased. We believe our work is interesting for future work on controllable pinning landscapes and flux focusing for applications. We believe our work suggests potential ways by which we can fabricate arrays of blind holes or pinning centers with comparable characteristics of mesoscopic dimensions in pure weakly pinned crystals where flux can be locally concentrated. These results are relevant not only for blind hole arrays, but also one could possibly explore similar behaviour in systems with different artificial pinning centers, like dopants, nano−rods etc., where the superconducting medium is singly connected and is present above or below the pins, as is the case for blind holes. Such studies would not only help in verifying the significance of the peculiarities of these pinning centers in inducing metastability in these systems, but also open up further possibilities for generation of controllable pinning landscapes and flux focusing techniques for applications.

**Acknowledgements**

SSB acknowledges help from Dr. Amit Banerjee and funding support from IIT Kanpur for the MOI facility setup in IIT Kanpur where all the experiments have been performed and DST Indo Japan and Spain (reference ACI 2009-0905) joint programs.